\documentclass[useAMS,usenatbib]{mn2e}
\usepackage{graphicx}
\usepackage{subfigure}
\usepackage{amssymb}
\usepackage{longtable}
\usepackage{multicol}
\usepackage{multirow}
\usepackage{lscape}

\title[Short Period CVs below the Period Gap]{Cataclysmic Variables below the Period Gap: 
Mass Determinations of 14 Eclipsing Systems}
\author[C. D. J. Savoury et al.]
{C. D. J. Savoury$^{1}$\thanks{E-mail:chris.savoury@sheffield.ac.uk}, S. P. Littlefair$^{1}$, 
V. S. Dhillon$^{1}$, T. R. Marsh$^{2}$, B. T. G\"{a}nsicke$^{2}$, \newauthor C. M. Copperwheat$^{2}$, 
P. Kerry$^{1}$, R. D. G. Hickman$^{2}$ and S. G. Parsons$^{2}$\\
$^{1}$Dept of Physics and Astronomy, University of Sheffield, Sheffield, S3 7RH, UK\\ 
$^{2}$Dept of Physics, University of Warwick, Coventry, CV4 7AL, UK}
\begin{document}

\date{Submitted for publication in the Monthly Notices of the Royal Astronomical 
Society \today}

\pagerange{\pageref{firstpage}--\pageref{lastpage}}  
\pubyear{2010}

\maketitle

\label{firstpage}

\begin{abstract}

We present high-speed, three-colour photometry of the eclipsing cataclysmic variables CTCV J1300-3052, 
CTCV J2354-4700 and SDSS J115207.00+404947.8. These systems have orbital periods of 128.07, 94.39 and 97.52 
minutes respectively, placing all three systems below the observed ``period gap'' for cataclysmic variables.
For each system we determine the system parameters by fitting a parameterised model to the observed eclipse
light curve by $\chi^{2}$ minimisation. 

We also present an updated analysis of all other eclipsing systems previously analysed by our group. The 
updated analysis utilises Markov Chain Monte Carlo techniques which enable us to arrive confidently at the 
best fits for each system with more robust determinations of our errors. A new bright spot model is also 
adopted, that allows better modelling of bright-spot dominated systems. In addition, we correct a bug in 
the old code which resulted in the white dwarf radius being underestimated, and consequently both the white 
dwarf and donor mass being overestimated. New donor masses are generally between $1$ and $2\sigma$ of those 
originally published, with the exception of SDSS 1502 ($-2.9\sigma$, $\Delta$$M_{r}=-0.012M_{\odot}$) and DV UMa 
($+6.1\sigma$, $\Delta$$M_{r}=+0.039M_{\odot}$). We note that the donor mass of SDSS 1501 has been revised upwards 
by $0.024M_{\odot}$ ($+1.9\sigma$). This system was previously identified as having evolved passed the minimum 
orbital period for cataclysmic variables, but the new mass determination suggests otherwise. Our new analysis 
confirms that SDSS 1035 and SDSS 1433 have evolved past the period minimum for cataclysmic variables, 
corroborating our earlier studies.

We find that the radii of donor stars are oversized when compared to theoretical models, by approximately
10 percent. We show that this can be explained by invoking either enhanced angular momentum loss, or by 
taking into account the effects of star spots. We are unable to favour one cause over the other, as we 
lack enough precise mass determinations for systems with orbital periods between 100 and 130 minutes, where 
evolutionary tracks begin to diverge significantly.

We also find a strong tendency towards high white dwarf masses within our sample, and no evidence for any 
He-core white dwarfs. The dominance of high mass white dwarfs implies that erosion of the white dwarf during 
the nova outburst must be negligible, or that not all of the mass accreted is ejected during nova cycles, 
resulting in the white dwarf growing in mass. 

\end{abstract}


\section{Introduction}
\label{sec:intro}
Cataclysmic variable stars (CVs) are a class of interacting binary system undergoing mass transfer from a Roche-lobe 
filling secondary to a white dwarf primary, usually via a gas stream and accretion disc. A bright spot is formed where 
the gas stream collides with the edge of the accretion disc, often resulting in an \textquoteleft orbital 
hump\textquoteright~in the light curve at phases $0.6-1.0$ due to the area of enhanced emission rotating into our line 
of sight. For an excellent overview of CVs, see \citet{warner1995a} and \citet{hellier2001}. The light curves of eclipsing 
CVs can be quite complex, with the accretion disc, white dwarf and bright spot all being eclipsed in rapid succession. 
When observed with time resolutions of the order of a few seconds, this eclipse structure allows the system parameters 
to be determined to a high degree of precision with relatively few assumptions \citep{wood1986}. Over the last eight 
years our group has used the high-speed, three-colour camera {\sc ultracam} \citep{dhillon2007} to obtain such 
time-resolution. The ability to image in three different wave-bands simultaneously makes {\sc ultracam} an ideal tool to study the 
complex, highly variable light curves of CVs. Using {\sc ultracam} data, we have obtained system parameters for several short 
period systems (e.g. Feline et al. 2004a, 2004b; Littlefair et al. 2006a, 2007, 2008), \nocite{feline2004a, feline2004a, 
feline2004b, littlefair2006a, littlefair2007, littlefair2008} including the first system accreting from a sub-stellar donor 
(Littlefair et al. 2006b). \nocite{littlefair2006b} 

Despite extensive study over recent decades, there are still several outstanding issues with evolutionary theories 
of CVs that have wide ranging implications for all close binary systems. The secular evolution of CVs is driven by 
angular momentum losses from the binary orbit. In the standard model, systems with orbital periods below $\sim$130 
minutes are thought to lose angular momentum via gravitational radiation. Angular momentum losses sustain mass transfer 
and subsequently drive the system to shorter orbital periods, until the point where the donor star becomes degenerate 
\citep[e.g.][]{paczynski1981}. Here, the donor star is driven out of thermal equilibrium and begins to expand in response 
to mass-loss, driving the system to longer orbital periods. We therefore expect to observe a period cut-off around 
$P_{orb} \simeq 65-70$ minutes, dubbed the ``period minimum'', in addition to a build up of systems at this minimum period 
(the ``period spike''; Kolb \& Baraffe 1999).\nocite{kolb1999} The period spike has recently been identified by 
\citet{gaensicke2009}, whose 
study of SDSS CVs found an accumulation of systems with orbital periods between 80 and 86 minutes. This is significantly 
longer than expected. A larger than expected orbital period implies that the orbital separation is larger than expected, 
nd thus the radii of the donor star must also be larger than expected in order to remain Roche-lobe filling. Recent 
observations by \citet{littlefair2008} support this, suggesting that the donor stars in short period CVs are roughly 
10 percent larger than predicted by the models of \citet{kolb1999}. The reason why the donor stars appear oversized 
remains uncertain. Possible explanations include some form of enhanced angular momentum loss \citep[e.g.][]{patterson1998, 
kolb1999, willems2005} which would increase mass loss and drive the donor stars further from thermal equilibrium, 
or missing stellar physics in the form of magnetic activity coupled with the effects of rapid rotation \citep[e.g.][]{chabrier2007}. 

One way to determine why the donor stars appear oversized is to compare the shape of the observed donor mass-period 
relationship ($M_{2}-P_{orb}$), and by implication, mass-radius ($M_{2}-R_{2}$) relationship, to the models of 
\citet{kolb1999}, calculated with enhanced angular momentum loss or modified stellar physics. These models, in principle, 
make different predictions for the shape of the mass-period relationship and the position of the period minimum. Both the 
shape of the mass-period relationship and the position of the period minimum are dependent on the ratio 
$\kappa = \tau_{M}/\tau_{KH}$, where $\tau_{M}$ and $\tau_{KH}$ are the mass-loss and thermal timescales of the donor star 
respectively. Initially, $\kappa$$\gg$1, and the donor is able to contract in response to mass-loss. As the system evolves 
to shorter orbital periods both timescales increase, although the thermal timescale increases much faster than the mass-loss 
timescale. This results in $\kappa$ decreasing with orbital period. When the two timescales become comparable, the donor 
is unable to contract rapidly enough to maintain thermal equilibrium and becomes oversized for a given mass. Since donor 
expansion does not occur until the thermal and mass-loss timescales become comparable, if enhanced angular momentum loss 
is responsible for the oversized CV donors, the systems {\it immediately} below the period gap would not be expected to be 
far from thermal equilibrium. In contrast, star spots would inhibit the convective processes in all CV donors below the 
period gap (assuming of course that spot properties are similar at all masses). Models that include the effects of enhanced 
angular momentum loss and star spot coverage therefore begin to diverge significantly in $\dot{M}$ and $M_{2}$ at orbital 
periods of 100 minutes. We can distinguish between these models if we have a sample of CVs that covers a wide range of 
orbital periods and whose component masses and radii are known to a high degree of precision (e.g. $\sigma$$M_{2}$$\sim0.005M_{\odot}$). 
Unfortunately we lack enough precise mass-radii determinations for systems with orbital periods between 95 and 130 minutes. 
To overcome this shortage we observed eclipses of three CVs below the period gap: CTCV J1300-3052, CTCV J2354-4700 and SDSS 
J115207.00+404947.8 (hereafter CTCV 1300, CTCV 2354 and SDSS 1152).

CTCV 2354 and CTCV 1300 were discovered as part of the Cal\'{a}n-Tololo Survey follow up \citep{tappert2004}. During the 
follow up, both systems were found to be eclipsing with orbital periods of 94.4 and 128.1 minutes, respectively. 
Basic, non-time resolved, spectroscopic data was obtained for each system. The spectrum of CTCV 1300 showed features typical 
of the three main components in CVs: strong emission lines from the accretion disc, broad, shallow absorption features from 
the white dwarf and red continuum and absorption bands from the donor. CTCV 2354 was found to contain strong emission lines 
of H and He, generally typical of a dwarf nova in quiescence. 

SDSS 1152 was identified as a CV by \citet{szkody2007}. The system shows broad, double-peaked, Balmer emission 
lines, which are characteristic of a high-inclination accreting binary. Follow up work by \citet{southworth2010} found the system 
to have an orbital period of 97.5 minutes.

In this paper we present {\sc ultracam} light curves of CTCV 1300 ($u'g'r'i'$), CTCV 2354 ($u'g'r'$) and SDSS 1152 ($u'g'r'$), 
and in each case attempt to determine the system parameters via light curve modelling. In addition, we also present an updated 
analysis of all eclipsing systems previously published by our group: OU Vir (Feline et al. 2004a), 
\nocite{feline2004a} XZ Eri and DV UMa (Feline et al. 2004b), \nocite{feline2004b} SDSS J1702+3229 (Littlefair et al. 2006a), 
\nocite{littlefair2006a} SDSS J1035+0551 (Littlefair et al. 2006b, 2008), \nocite{littlefair2006b, littlefair2008} 
SDSS J150722+523039 \citep{littlefair2007, littlefair2008}, SDSS J0903+3300, SDSS J1227+5139, SDSS J1433+1011, SDSS J1501+5501 
and SDSS J1502+3334 \citep{littlefair2008}. Our primary reason for doing so was the introduction of a new analysis utilising 
Markov Chain Monte Carlo (MCMC) techniques and an updated bright spot model. The MCMC analysis is more reliable at converging to a best 
fit than the downhill simplex algorithm used previously, while the new bright spot model allows for a more realistic modelling of 
bright spot dominated systems (e.g. CTCV 1300, DV UMa, SDSS 1702) and should thus provide more accurate values of the mass ratio, $q$. 
While implementing these changes, we also discovered a bug in the code previously used to bin the light curves (see e.g. 
Littlefair et al. 2006a \nocite{littlefair2006a} for details of the original code) which resulted in the white 
dwarf radius being underestimated, and consequently, the white dwarf and donor mass being overestimated. Full details 
are provided in section \ref{sec:pme}.     

\section{Observations}
\label{sec:obs}

In Table \ref{table:journal} we present details of the observations used to analyse CTCV 1300, CTCV 2354, SDSS 1152 and
SDSS 1501. For observations
of other systems we refer the reader to table 1 in the following publications: Feline et al. 2004a (OU Vir), Feline et al. 
2004b (XZ Eri and DV UMa), Littlefair et al. 2006a (SDSS 1702), Littlefair et al. 2007 (SDSS 1507) and Littlefair et al. 2008 
(SDSS 0903, SDSS 1227, SDSS 1433, SDSS 1501 and SDSS 1502). 

For reasons outlined in section \ref{sec:pme} we do not use the SDSS 1501 data listed in Littlefair et al. (2008). Instead we model 
a single eclipse observed in 2004 (Table \ref{table:journal}). Not all of the eclipses listed in Table \ref{table:journal} 
are used for determining system parameters. This is because the eclipses have poor signal-to-noise, or lack clear bright spot 
features. The eclipses not used for determining system parameters are however still used to refine 
our orbital ephemerides (section 
\ref{sec:ephem}). These eclipses include CTCV 2354 cycle numbers 11197, 11198, 11366, 11396, 11457, 11472 and SDSS 1501 cycles 
24718 and 24719. CTCV 2354 cycle numbers 16156, 16676 and CTCV 1300 cycle number 12888 are analysed separately in section 
\ref{sec:notes} because the shape of the eclipse has changed significantly in comparison to the 2007 data (see section \ref{sec:lcm}).

Data reduction was carried out in a standard manner using the {\sc ultracam} pipeline reduction software, as described in 
\citet{feline2005} and \citet{dhillon2007}. A nearby comparison star was used to correct the data for transparency variations. 
Observations of the standard stars G162-66, G27-45 and G93-48 were used to correct the magnitudes to the standard SDSS system 
\citep{smith2002}. Due to time constraints and poor weather, we were unable to observe a standard star to flux calibrate our 
data for SDSS 1152. Consequently, we have used the Sloan magnitudes of the comparison stars and corrected for different 
instrumental response. To do this, we use measured response curves for filters and dichroics to create overall response 
curves for {\sc ultracam}. These are then combined with curves of theoretical extinction and library spectra 
\citep{pickles1998} to obtain synthetic {\sc ultracam} colours. The same process is then repeated for the SDSS colour set,
with the difference between the two sets being the correction applied.

\begin{table*}
\caption{Journal of observations. The dead-time between exposures was 0.025~s for all observations. 
The relative GPS time stamping on each data point is accurate to 50 $\mu$s. Instr setup denotes the telescope
(WHT, NTT or VLT) and instrument used for each observation, where UCAM and USPEC represent {\sc ultracam} and 
{\sc ultraspec}, respectively. Phase Cov corresponds to the phase coverage of the eclipse, taking the eclipse of the
white dwarf as phase 1. $T_{exp}$ and $N_{exp}$ denote the exposure time, and number of exposures, respectively. }
\centering
\begin{tabular}{ccccrccrcc}
\hline
\hline
Date & Object & Instr setup & $T_{mid}$ (HMDJ) & Cycle & Phase Cov & Filters &  $T_{exp}$ (s) & $N_{exp}$ & Seeing ('')\\ 
\hline
2007 June 09 & CTCV 2354 & VLT+UCAM  & 54261.383926(25) &     0 & 0.73--1.08 & $u'g'r'$ & 2.22 &  821 & 0.6--1.0\\ 
2007 June 13 & CTCV 2354 & VLT+UCAM  & 54265.316786(61) &    60 & 0.70--1.11 & $u'g'r'$ & 4.92 &  473 & 0.8--1.2\\
2007 June 15 & CTCV 2354 & VLT+UCAM  & 54267.348921(21) &    91 & 0.78--1.08 & $u'g'r'$ & 2.22 &  779 & 0.6--0.7\\
2007 June 15 & CTCV 2354 & VLT+UCAM  & 54267.414476(20) &    92 & 0.74--1.05 & $u'g'r'$ & 2.22 &  757 & 0.6--0.7\\
2007 June 16 & CTCV 2354 & VLT+UCAM  & 54268.397717(20) &   107 & 0.72--1.06 & $u'g'r'$ & 2.22 &  845 & 0.6--1.1\\
2007 June 19 & CTCV 2354 & VLT+UCAM  & 54271.413077(21) &   153 & 0.82--1.15 & $u'g'r'$ & 2.22 &  826 & 0.6--1.0\\
2007 June 20 & CTCV 2354 & VLT+UCAM  & 54272.396368(29) &   168 & 0.50--1.50 & $u'g'r'$ & 2.32 & 2390 & 0.6--1.0\\
2007 June 21 & CTCV 2354 & VLT+UCAM  & 54273.314054(5)  &   182 & 0.86--1.20 & $u'g'r'$ & 1.96 &  931 & 1.2--2.4\\
2007 June 21 & CTCV 2354 & VLT+UCAM  & 54273.379579(3)  &   183 & 0.77--1.09 & $u'g'r'$ & 1.96 &  916 & 0.9--1.5\\
2009 June 12 & CTCV 2354 & NTT+USPEC & 54995.350263(6)  & 11197 & 0.65--1.35 & $ g'   $ & 9.87 &  482 & 1.4--2.6\\
2009 June 12 & CTCV 2354 & NTT+USPEC & 54995.415961(6)  & 11198 & 0.35--1.17 & $ g'   $ & 9.87 &  482 & 1.0--2.2\\
2009 June 23 & CTCV 2354 & NTT+USPEC & 55006.428224(2)  & 11366 & 0.66--1.16 & $ g'   $ & 3.36 &  817 & 1.4--2.2\\
2009 June 25 & CTCV 2354 & NTT+USPEC & 55008.394766(1)  & 11396 & 0.70--1.22 & $ g'   $ & 3.36 &  855 & 1.2--2.0\\
2009 June 29 & CTCV 2354 & NTT+USPEC & 55012.393334(1)  & 11457 & 0.55--1.37 & $ g'   $ & 2.98 & 1593 & 0.8--2.4\\
2009 June 30 & CTCV 2354 & NTT+USPEC & 55013.376588(1)  & 11472 & 0.30--1.55 & $ g'   $ & 1.96 & 3592 & 1.2--2.0\\
2010 May~ 03 & CTCV 2354 & NTT+UCAM  & 55320.414(2)     & 16156 & 0.43--1.21 & $u'g'r'$ & 8.23 &  526 & 1.4--1.8\\
2010 June 06 & CTCV 2354 & NTT+UCAM  & 55354.434846(24) & 16676 & 0.47--1.19 & $u'g'r'$ & 3.84 & 1048 & 1.0--1.2\\
\hline
2007 June 10 & CTCV 1300 & VLT+UCAM  & 54262.099145(3)  &     0 & 0.72--1.20 & $u'g'r'$ & 1.00 & 3462 & 0.6--1.2\\
2007 June 13 & CTCV 1300 & VLT+UCAM  & 54262.123093(8)  &    34 & 0.74--1.15 & $u'g'i'$ & 1.95 & 1573 & 0.6--1.1\\
2010 June 07 & CTCV 1300 & NTT+UCAM  & 55355.002677(1)  & 12288 & 0.85--1.12 & $u'g'r'$ & 2.70 &  511 & 0.8--1.1\\
\hline 
2010 Jan~ 07 & SDSS 1152 & WHT+UCAM  & 55204.101282(9)  &     0 & 0.16--1.13 & $u'g'r'$ & 3.80 & 1492 & 2.0--3.8\\
2010 Jan~ 07 & SDSS 1152 & WHT+UCAM  & 55204.169031(8)  &     1 & 0.72--1.12 & $u'g'r'$ & 3.80 &  600 & 1.4--2.5\\
2010 Jan~ 07 & SDSS 1152 & WHT+UCAM  & 55204.236742(7)  &     2 & 0.85--1.12 & $u'g'r'$ & 3.80 &  415 & 1.2--3.2\\
\hline
2004 May~ 17 & SDSS 1501 & WHT+UCAM  & 53142.921635(6)  &-11546 & 0.80--1.21 & $u'g'r'$ & 6.11 &  335 & 1.0--1.6\\
2010 Jan~ 07 & SDSS 1501 & WHT+UCAM  & 55204.213149(3)  & 24718 & 0.78--1.12 & $u'g'r'$ & 3.97 &  435 & 1.4--4.0\\
2010 Jan~ 07 & SDSS 1501 & WHT+UCAM  & 55204.270000(3)  & 24719 & 0.88--1.13 & $u'g'r'$ & 3.97 &  321 & 1.4--3.0\\
\hline
\hline
\end{tabular}
\label{table:journal}
\end{table*}

\section{Results}

\subsection{Orbital ephemerides}
\label{sec:ephem}
The times of white dwarf mid-ingress $T_{wi}$ and mid-egress $T_{we}$ were determined by locating the minimum and 
maximum times, respectively, of the smoothed light-curve derivative. Mid-eclipse times, $T_{mid}$, were determined 
by assuming the white dwarf eclipse to be symmetric around phase one and taking $T_{mid}=(T_{wi}+T_{we})/2$. Eclipse
times were taken from the literature for CTCV 1300, CTCV 2354 \citep{tappert2004}, SDSS 1501 \citep{littlefair2008}
and SDSS 1152 \citep{southworth2010} and combined with our mid-eclipse times shown in Table \ref{table:journal}. The 
errors on our data were adjusted to give $\chi^{2} = 1$ with respect to a linear fit. In each case we observe no cycle 
ambiguity. We do however, observe a significant, O-C offset between our data and the times published by \citet{tappert2004} 
for CTCV 1300 and CTCV 2354. For CTCV 1300, the average difference is 165.9 seconds, while for CTCV 2354 it is 
148.0 seconds. We believe this to be due to the differing methods of calculating $T_{mid}$; Tappert et al. (2004) calculated  
$T_{mid}$ by fitting a parabola to the overall eclipse structure, whereas we determined $T_{mid}$ from the white dwarf 
eclipse. We therefore subtract these average offsets from the published literature values and take the resulting O-C 
difference as our uncertainty on that time. Where possible, we averaged the measured mid-eclipse times in the $r'$ and 
$g'$ bands to fit the ephemeris. Due to the low signal-to-noise of the eclipse of CTCV 2354 in May 2010, measuring the 
times of white dwarf mid-ingress $T_{wi}$ and mid-egress $T_{we}$ was not possible. Consequently, the eclipse times and 
errors were measured by eye in the $g'$ and $r'$ bands and then averaged in order to estimate the cycle 
number. This time was not used to refine the ephemeris but is included for completeness. The ephemerides found are shown 
in Table \ref{table:ephemeris}. 

\begin{table}
\caption{Orbital ephemerides.}
\centering
\begin{tabular}{ccc}
\hline
Object & $T_{\rmn{0}}$ (HJD) & $P_{\rmn{orb}}$ (d)\\
\hline
CTCV 1300 & 2454262.599146 (8) & 0.088940717 (1) \\
CTCV 2354 & 2454261.883885 (5) & 0.065550270 (1) \\
SDSS 1152 & 2455204.601298 (6) & 0.067721356 (3) \\
SDSS 1501 & 2453799.710832 (3) & 0.0568412623(2) \\
\hline
\end{tabular}
\label{table:ephemeris}
\end{table}

\subsection{Light curve morphology and variations}
\label{sec:lcm}
\subsubsection{CTCV J1300-3052}
Fig. \ref{figure:lightcurves} (top) shows the two observed eclipses of CTCV 1300 from the 2007 data 
set folded on orbital phase in the $g'$ band. The white dwarf ingress and egress features are clearly 
visible at phases 0.965 and 1.040, respectively, as are the bright spot features at phases 0.960 and 
1.085. These features dominate the light curve, which follows a 
typical dwarf nova eclipse shape (e.g. Littlefair et al. 2006a, 2007, 2008). The depth of the 
bright spot eclipse indicates that the bright spot is the dominant source of light in this system, 
while the eclipse of the accretion disc is difficult to discern by eye, indicating that the accretion 
disc contributes little light to this system. Closer inspection of the eclipses from each night reveals 
a noticeable difference in the shape of the bright spot ingress feature. This is caused by heavy pre-eclipse
flickering, and is clearly visible in Fig. \ref{fig:eclipses}. The flickering is reduced between 
phases corresponding to the white dwarf ingress and bright spot egress, indicating the source of 
the flickering is the inner disc and/or bright spot. Due to the heavy flickering, we decided to fit 
each night individually rather than fit to a phase-folded average, in order to provide a more robust 
estimation of our uncertainties. Our 2010 observations are discussed in section \ref{sec:ctcv1300_2010}.

\subsubsection{CTCV J2354-4700}
\label{sec:2354}
Fig. \ref{figure:lightcurves} (middle) shows all of the observed eclipses from the 2007 data set of 
CTCV 2354 folded on orbital phase in the $g'$ band. The white dwarf ingress and egress features are 
clearly visible and along with the accretion disc dominate the shape of the light curve. A weak bright 
spot ingress feature is visible at an orbital phase of 0.995, however the system suffers from heavy 
flickering, making it difficult to identify the bright spot egress. The shape of the average light curve 
indicates possible egress features at phases 1.060 and 1.080, but given the scatter we cannot be certain 
whether these represent genuine egress features or merely heavy flickering. The flickering is reduced between 
phases corresponding to the white dwarf ingress and egress, indicating the source of the flickering is 
the inner disc. Our observations from 2009 and 2010 are discussed in section \ref{sec:ctcv2354_2010}.

\subsubsection{SDSS J1152+4049}
\begin{figure}
\centering
\includegraphics[scale=0.40,angle=0,trim=0 0 0 0,clip]{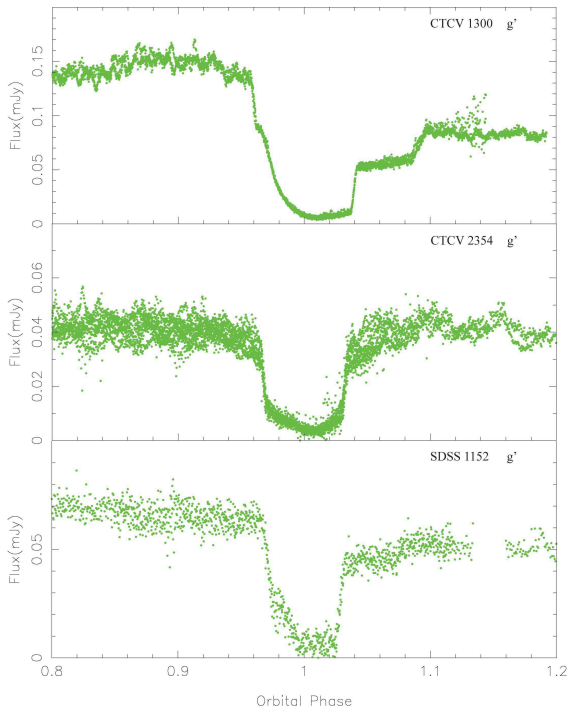}
\caption{{\sc ultracam} $g'$ band light curves of CTCV 1300 (2007, top), CTCV 2354 (2007, middle) and SDSS 1152 (2010, bottom).}
\label{figure:lightcurves}
\end{figure}

Fig. \ref{figure:lightcurves} (bottom) shows all of the observed eclipses from the 2010 data set of SDSS 1152 
folded on orbital phase in the $g'$ band. The signal-to-noise ratio of our data is low in comparison to other 
systems, but we still see a clear bright spot ingress feature at phase 0.975 in addition to a clear bright spot 
egress feature at phase 1.075. The white dwarf features are clear, and dominate the overall shape of the light 
curve. Like CTCV 1300, the eclipse of the accretion disc is difficult to discern by eye, which again suggests 
that the accretion disc contributes little light to this system. 

\subsection{Light curve modelling}
\label{sec:pme}

To determine the system parameters we used a physical model of the binary system to calculate eclipse light curves 
for the white dwarf, bright spot, accretion disc and donor. Feline et al. (2004b) \nocite{feline2004b} showed that 
this method gives a more robust determination of the system parameters in the presence of flickering than the derivative 
method of \citet{wood1986}. The model itself is based on the techniques developed by \citet{wood1985} and 
\citet{horne1994}, and is an adapted version of the one used by \citet{littlefair2008}. This model
relies on three critical assumptions: the bright spot lies on the ballistic trajectory from the donor 
star, the donor fills its Roche lobe, and the white dwarf is accurately described by a theoretical mass-radius 
relation. Obviously these assumptions cannot be tested directly, but it has been shown that the 
masses derived with this model are consistent with other methods commonly employed in CVs over 
a range of orbital periods \citep[e.g.][]{feline2004b, tulloch2009, copperwheat2010}. 
The model used by \citet{littlefair2008} had to be adapted due to the prominent
bright spot observed in CTCV 1300 (Fig. \ref{figure:lightcurves}). The \textquoteleft old\textquoteright~model 
fails to correctly model the bright spot ingress and egress features in this system satisfactorily and results in 
a poor fit. We have thus adapted the model to account for a more complex bright spot by adding four new parameters, 
following \citet{copperwheat2010}, bringing the total number of variables to 14. These are: 

\begin{enumerate}
\item The mass ratio, $q=M_r/M_w$.
\item The white dwarf eclipse phase full-width at half-depth, $\Delta\phi$.
\item The outer disc radius, $R_{d}/a$, where $a$ is the binary separation.
\item The white dwarf limb-darkening coefficient, $U_{w}$.
\item The white dwarf radius, $R_{w}/a$.
\item The bright-spot scale, $S/a$. The bright spot is modelled as two
linear strips passing through the intersection of the gas stream and
disc. One strip is isotropic, while the other beams in a given direction. 
Both strips occupy the same physical space. The intensity distribution is 
given by $(X/S)^{Y}e^{-(X/S)^{Z}}$, where $X$ is the distance along the strips.
The non-isotropic strip does not beam perpendicular to its surface. Instead 
the beaming direction is defined by two angles, $\theta_{tilt}$ and 
$\theta_{yaw}$.
\item The first exponent, $Y$, of the bright spot intensity distribution.
\item The second exponent, $Z$, of the bright spot intensity distribution.
\item The bright-spot angle, $\theta_{az}$, measured relative to the line 
joining the white dwarf and the secondary star. This allows adjustment of 
the phase of the orbital hump.
\item The tilt angle, $\theta_{tilt}$, that defines the beaming direction of 
the non-isotropic strip. This angle is measured out of the plane of 
the disc, such that $\theta_{tilt} = 0$ would beam light perpendicular to the 
plane of the disc.
\item The yaw angle, $\theta_{yaw}$. This angle also defines the beaming direction 
of the non-isotropic strip, but in the plane of the disc and with respect to the 
first strip. 
\item The fraction of bright spot light that is isotropic, $f_{iso}$.
\item The disc exponent, $b$, describing the power law of the radial
intensity distribution of the disc.
\item A phase offset, $\phi_{0}$.
\end{enumerate}

The data are not good enough to determine the white dwarf limb-darkening coefficient, $U_{w}$, 
accurately. To find an appropriate limb-darkening coefficient, we follow the procedure outlined in 
\citet{littlefair2007}, whereby an estimate of the white dwarf effective temperature and mass is 
obtained from a first iteration of the fitting process outlined below, assuming a limb-darkening 
coefficient of 0.345. \citet{littlefair2007} show that typical uncertainties in $U_{w}$ are 
$\sim5$\%, which leads to uncertainties in $R_{w}/a$ of $\sim1$\%. These errors have negligible 
impact on our final system parameters. 

As well as the parameters described above, the model also provides an estimate of the flux contribution 
from the white dwarf, bright spot, accretion disc and donor. The white dwarf temperature and distance are 
found by fitting the white dwarf fluxes from our model to the predictions of white dwarf model atmospheres 
\citep{bergeron1995}, as shown in Fig. \ref{fig:wd_colours}. We find that with the exception of 
CTCV 2354, all of the systems analysed lie near, or within, the range of white dwarf colours allowed 
by the atmosphere models of \citet{bergeron1995}, although the systems do not always lie near the track 
for the appropriate mass and radius of the white dwarf. \citet{littlefair2008} compare the temperatures 
derived using light curve fits to those found using SDSS spectra and GALEX (Galaxy Evolution Explorer) 
fluxes for a small number of systems and conclude their white dwarf temperatures are accurate to $\sim$1000K. 
The systems examined by Littlefair et al. (2008) are all found to lie close to the Bergeron tracks; it 
is likely that systems that lie far from the tracks are less accurate. We note our temperatures
have larger uncertainties than those of \citet{littlefair2008}. This is because our temperatures 
take into account the uncertainty in white dwarf mass when comparing the white dwarf fluxes to the
models of \citet{bergeron1995}.

It is possible that our white dwarf colours are affected by contamination from the disc or bright spot,
or an unmodelled light source such as a boundary layer. If our white dwarf colours are incorrect, then 
our derived white dwarf temperatures will be affected. Changing the white dwarf temperature will alter 
$U_{w}$. Our model fitting measures $R_{w}/a$ and uses a mass-radius relationship to infer $M_{w}$, which 
is then used to find the mass of the donor star. However, $U_{w}$ and $R_{w}$ are partially degenerate, 
so $U_{w}$ therefore affects $R_{w}$ and $M_{w}$. $M_{w}$ is also affected by temperature changes because 
the white dwarf mass-radius relationship is temperature dependent. The white dwarf temperature also affects 
the luminosity of the system, and hence distance estimate. It is therefore important to quantify the effect 
that incorrect white dwarf temperatures may have on distance estimates and our final derived system 
parameters. To do this, we altered the white dwarf temperature by 2000K and performed the fitting procedure 
described above on our best quality, white-dwarf dominated systems. For lower quality data, the random 
errors dominate over any systematic errors, and thus changes to the best quality data represent a worst 
case scenario. We find that changing the white dwarf temperature by 2000K changes $R_{w}/a$ by 
less than 1$\sigma$. The white dwarf distance estimates change by 10-20pc. We therefore conclude any error 
in white dwarf temperature that may occur does not affect our final system parameters by a significant amount.
We note here that our moedelling does not include treatment of any boundary layer around the white dwarf, 
and assumes all of the white dwarf's surface is visible. Either effect could lead to systematic uncertainty
in our white dwarf radii \citep{wood1986}.

A Markov Chain Monte Carlo (MCMC) analysis was used to adjust all parameters bar $U_W$. 
MCMC analysis is an ideal tool as not only does it provide a robust method for quantifying 
the uncertainties in the various system parameters, it is more likely to converge on the global minimum 
$\chi^{2}$ rather than a local minimum $\chi^{2}$. We refer the reader to \citet[][]{ford2006}, 
\citet[][]{gregory2007} and references therein for excellent overviews of MCMC chains and Bayesian 
statistics and limit ourselves to a simple overview. 

MCMC is a random walk process where at each step in the chain we draw a set of model parameters from a normal, 
multi-variate distribution. This is governed by a covariance array, which we estimate from the initial stages 
of the MCMC chain. The step is either accepted or rejected based on a transition probability, which is a function 
of the change in $\chi^{2}$. We adopt a transition probability given by the Metropolis-Hastings (M-H) rule, that 
is $P = \exp^{-\Delta\chi^{2}/2}$. The sizes of the steps in the MCMC chain are multiplied by a scale factor, tuned 
to keep the acceptance rate near 0.23, which is found to be the optimal value for multi-variate chains such as 
these \citep{roberts1997}.

A typical MCMC chain included some 700,000 steps, split into two, 350,000 step sections. The first section is used to
converge {\it towards} the global minimum and estimate the covariance matrix (known as the burn-in phase). 
The second section fine tunes the solution by sampling areas of parameter space around the minimum. In doing so, 
we also produce a robust estimation of our uncertainties. Together, these steps are usually sufficient to enable the 
model to converge on the statistical best fit, regardless of the initial starting parameters. 

While implementing the MCMC code, we discovered a bug in our original code. The re-binning code used to average 
several light curves together mistreated the widths of the bins, which in turn affected the trapezoidal integration 
of the model over these bins. The direct result was that in cases of heavy binning, such as systems with heavy flickering 
or where several light curves had been averaged together (e.g. SDSS 1502), the white dwarf radius, $R_{w}/a$, was underestimated. 
The exact amount depended on the level of binning used. This consequently resulted in an overestimate of the white dwarf 
mass. Since the mass of the donor star, $M_{r}$, is related to the white dwarf mass $M_{w}$ by $M_{r} = qM_{w}$, we were 
also left with an overestimate of the donor mass. This problem affects all of our previously published eclipsing-CV papers 
(Feline et al. 2004a, 2004b; Littlefair et al. 2006a, 2006b, 2007, 2008) \nocite{feline2004a, feline2004b, littlefair2006a, 
littlefair2006b, littlefair2007, littlefair2008} by differing amounts. However, in most cases re-modelling provides new system 
parameters that are within $1-2 \sigma$ of our original results, with only two exceptions (see section \ref{sec:notes}). 
The new results are presented in Table \ref{table:system_params}.

For each system we ran an MCMC simulation on each phase-folded $u'$, $g'$, $r'$ or $i'$ light curve from an arbitrary 
starting position. Exceptions include CTCV 1300, where each night of observations was fit individually, and SDSS 1152 
and SDSS 1501, for which we only calculated fits in the $g'$ and $r'$ bands due to $u'$-band data of insufficient quality 
to constrain the model. Where no $u'$ band MCMC fit could be obtained, we fit and scaled the $g'$ band model to the $u'$ 
band light curves without $\chi^{2}$ optimisation. This allows us to estimate the white dwarf flux in the $u'$ band, and 
thus estimate the white dwarf temperature. In the case of SDSS 1501, we also fit a different data set to the 2006 WHT data 
of \citet{littlefair2008}. We fit 
our model to the single light curve dated 2004 May 17. This 2004 data was not fit by Littlefair et al. (2008) as the simplex 
methods used gave a seemingly good fit to the 2006 data. Despite appearing to have converged to a good fit, the MCMC 
analysis revealed that the 2006 data does not constrain the model, most likely due to the very weak bright spot features.
The 2004 data shows much clearer and well-defined bright-spot features than the 2006 data (see Fig.1 of Littlefair et al.
2008), and so despite only having one eclipse (and thus lower signal-to-noise) it is favoured for the fitting process. 
In general our fits to each system are in excellent agreement with the light curves (see Fig. \ref{fig:eclipses}), 
giving us confidence that our new models accurately describe each system. 

To obtain final system parameters we combine our MCMC chains with Kepler's $3^{rd}$ law, the orbital period, our 
derived white dwarf temperature, and a series of white dwarf mass-radius relationships. We favour the relationships 
of \citet{wood1995}, because they have thicker hydrogen layers which may be more appropriate for CVs.
However, they do not reach high enough masses for some of our systems. Above $M_{w} = 1.0 M_{\odot}$, 
we adopt the mass-radius relationships of \citet{panei2000}. In turn, these models do not extend beyond 
$M_{w} = 1.2 M_{\odot}$; above this mass we use the \citet{hamada1961} relationship. No attempt is made to
remove discontinuities from the resulting mass-radius relationship.

We calculate the mass ratio $q$, white dwarf mass $M_{w}/M_{\odot}$, white dwarf radius $R_{w}/R_{\odot}$, donor 
mass $M_{r}/M_{\odot}$, donor radius $R_{r}/R_{\odot}$, inclination $i$, binary separation $a/R_{\odot}$ and radial velocities
of the white dwarf and donor star ($K_{w}$ and $K_{r}$, respectively) for each step of the MCMC chain. Since each step of the 
MCMC has already been accepted or rejected based upon the Metropolis-Hastings rule, the distribution function for each parameter
gives an estimate of the probability density function (PDF) of that parameter, given the constraints of our eclipse data.
We can then combine the PDFs obtained in each band fit into the total PDF for each system, as shown in Fig. \ref{fig:pdfs}. 
We note that most systems have system parameters with a Gaussian distribution with very little asymmetry. Our adopted value 
for a given parameter is taken from the peak of the PDF. Upper and lower error bounds are derived from the 67\% confidence
levels. For simplicity, since the distributions are mostly symmetrical, we take an average of the upper and lower error bounds. 
The final adopted system parameters are shown in Table \ref{table:system_params}, although Fig. \ref{figure:wd_masses} and 
\ref{figure:models} show the true 67\% confidence levels for the white dwarf mass and donor mass respectively, for each system.

\begin{figure*}
\centering
\includegraphics[scale=0.80, trim=0 0 0 0]{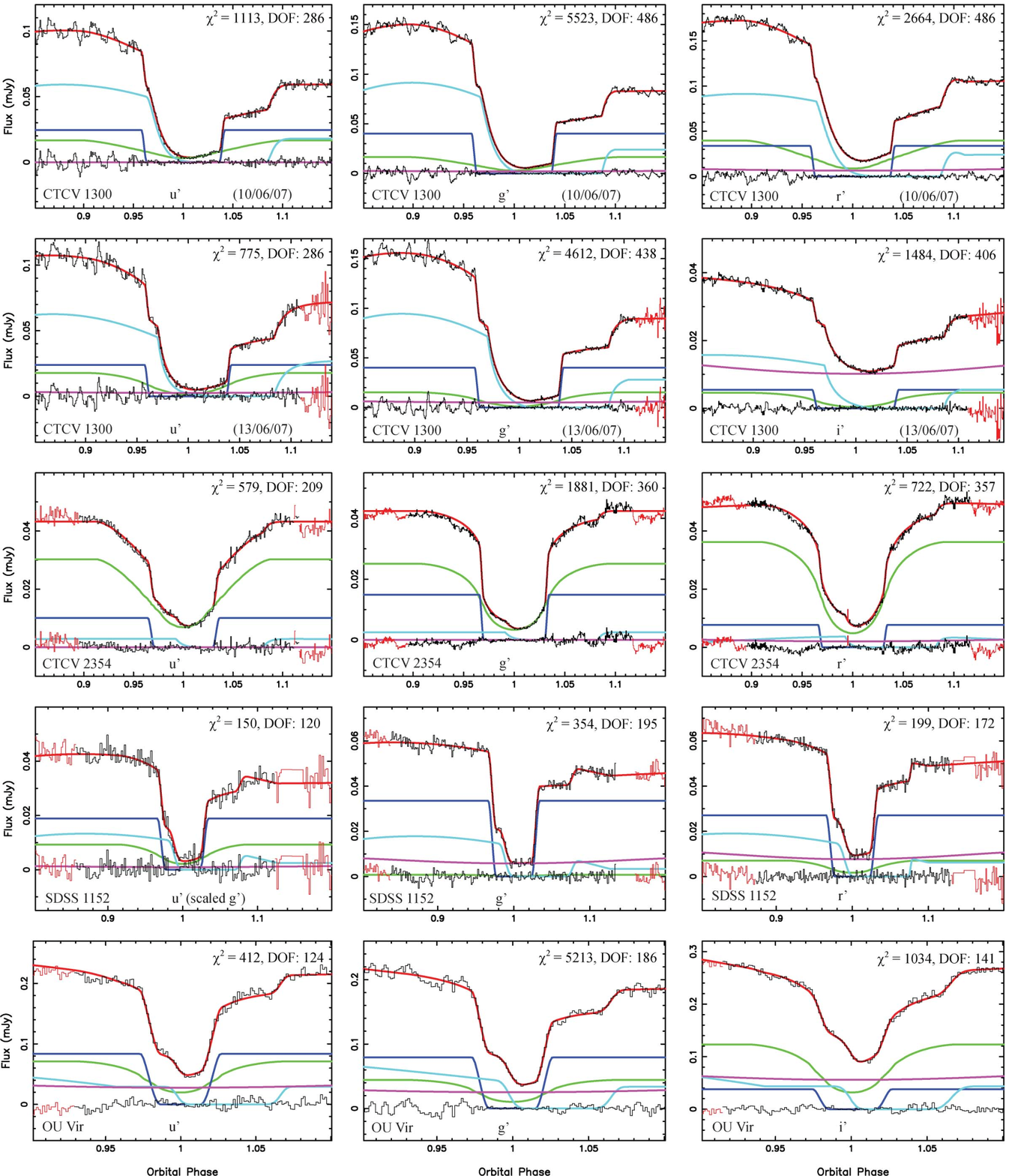}
\caption{The phased-folded $u'g'r'$ or $u'g'i'$ light curves of the CVs listed in Table \ref{table:system_params}, 
fitted using the model outlined in section \ref{sec:pme}. 
The data (black) are shown with the fit (red) overlaid and the residuals plotted below (black). Below are the 
separate light curves of the white dwarf (dark blue), bright spot (light blue), accretion disc (green) and the 
secondary star (purple). Data points omitted from the fit are shown in red. $\chi^{2}$ values for each fit, together
with the number of degrees of freedom (DOF) are also shown.}
\label{fig:eclipses}
\end{figure*}

\begin{figure*}
\begin{center}
\includegraphics[scale=0.80, trim=0 0 0 0]{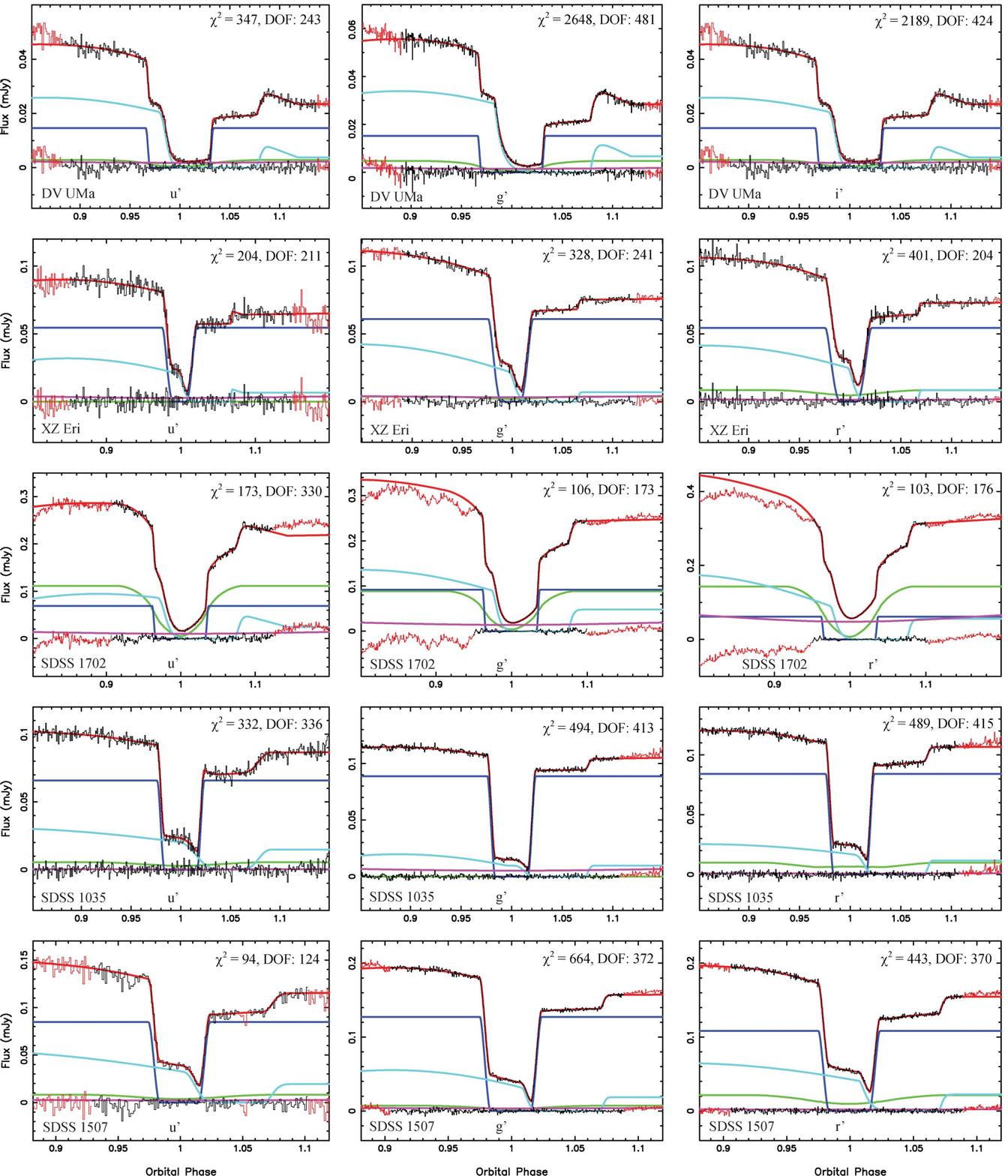}
\end{center}
\contcaption{}
\end{figure*}

\begin{figure*}
\begin{center}
\includegraphics[scale=0.80, trim=0 0 0 0]{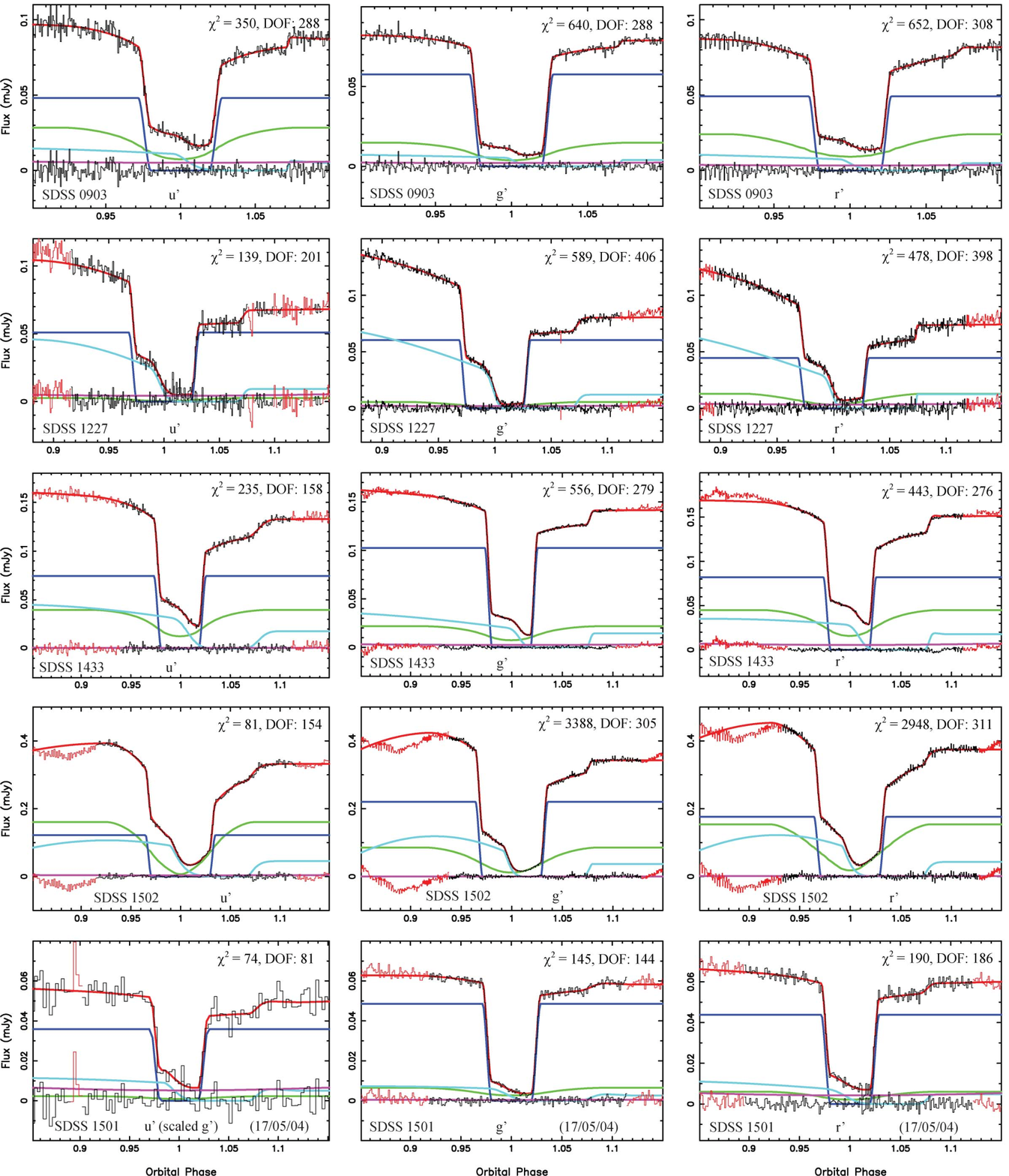}
\end{center}
\contcaption{}
\end{figure*}

\begin{figure*}
\centering
\includegraphics[scale=0.80, trim=0 0 0 0]{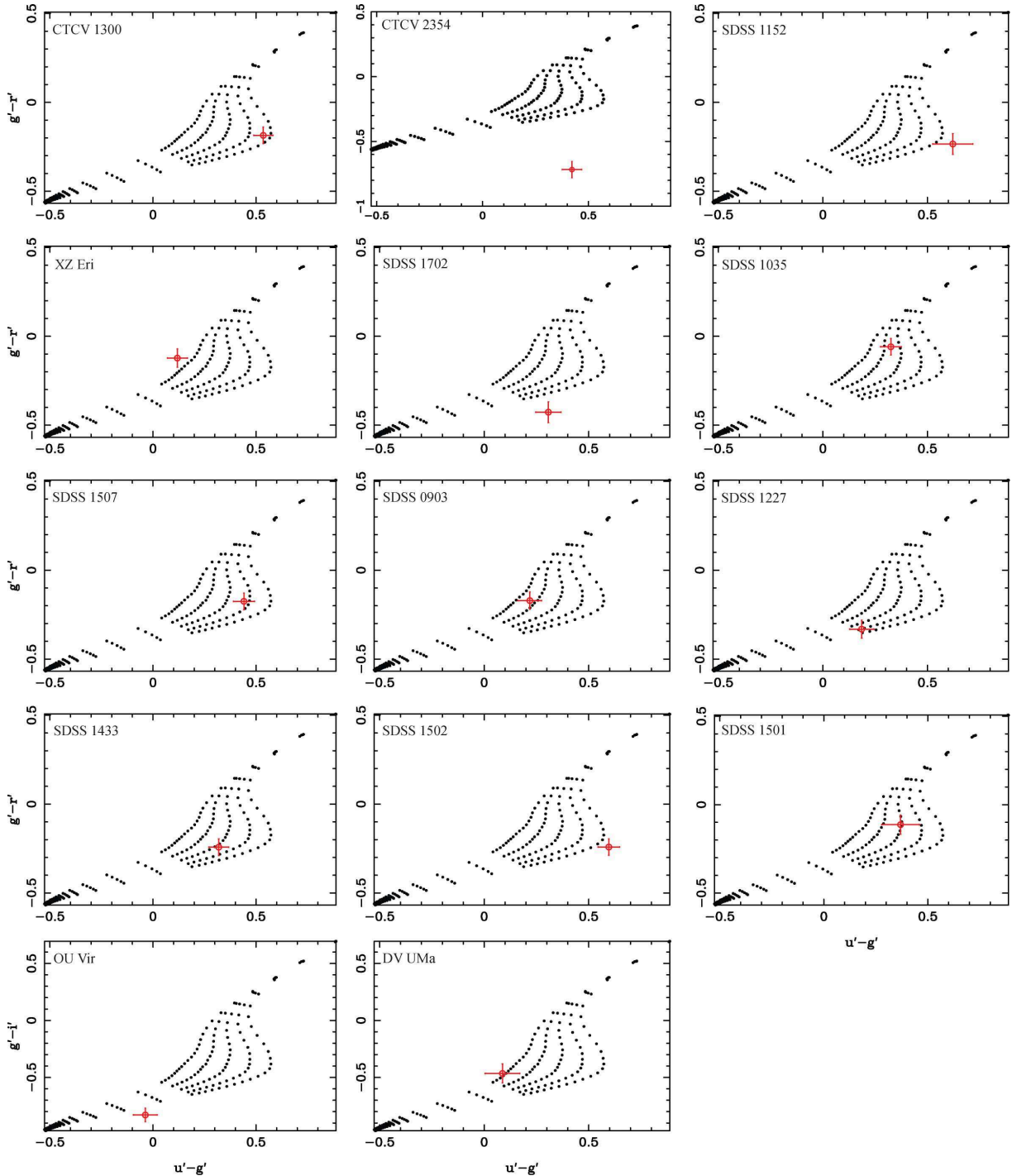}
\caption{The white dwarf colours derived from our model fitted together with the white dwarf models of \citet{bergeron1995}.
From top to bottom, each curve represents log g = 9.0, 8.5, 8.0, 7.5 and 7.0 respectively. The measured white dwarf colours 
are shown here in red, and are used to derive the white dwarf temperature, which in turn is used to correct the white dwarf 
mass-radius relationships used later to obtain the final system parameters.} 
\label{fig:wd_colours}
\end{figure*}

\begin{figure*}
\centering
\includegraphics[scale=0.80]{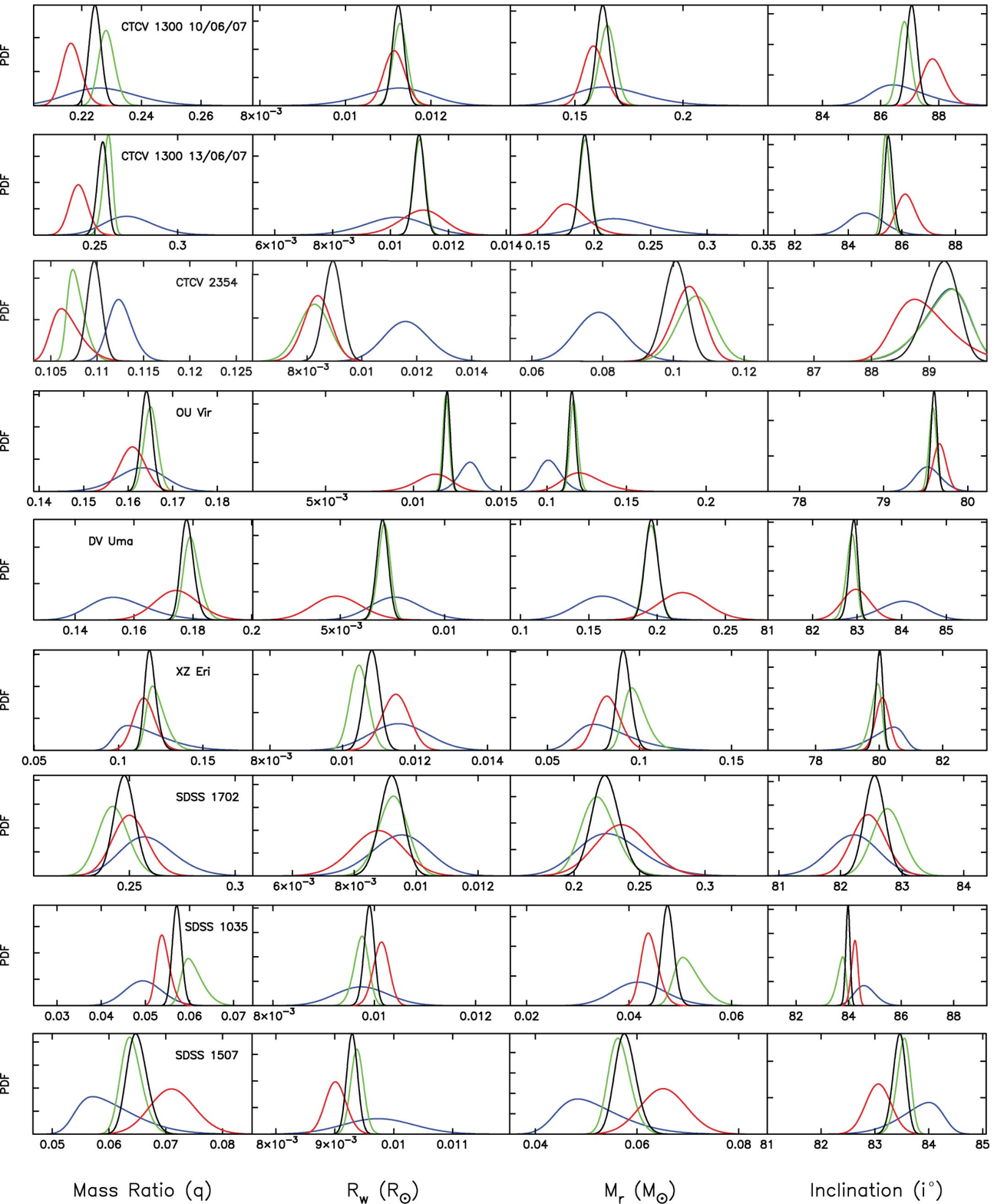}
\caption{The normalised probability density functions for each system, derived using the MCMC chains, orbital period 
and the mass--radius relationships of \citet{wood1995}, \citet{panei2000} and \citet{hamada1961}, at the appropriate 
white dwarf temperature. 
The red curve represents the $r'$ or $i'$ band fit, the green represents the $g'$ band, and blue curve (where present) represents 
the $u'$ band. The black represents the total, combined PDF. Shown are the PDFs for mass ratio $q$, white dwarf radius 
$R_{w}/R_{\odot}$, donor mass $M_{r}/M_{\odot}$, and inclination $i^{\circ}$.}
\label{fig:pdfs}
\end{figure*}

\begin{figure*}
\begin{center}
\includegraphics[scale=0.80]{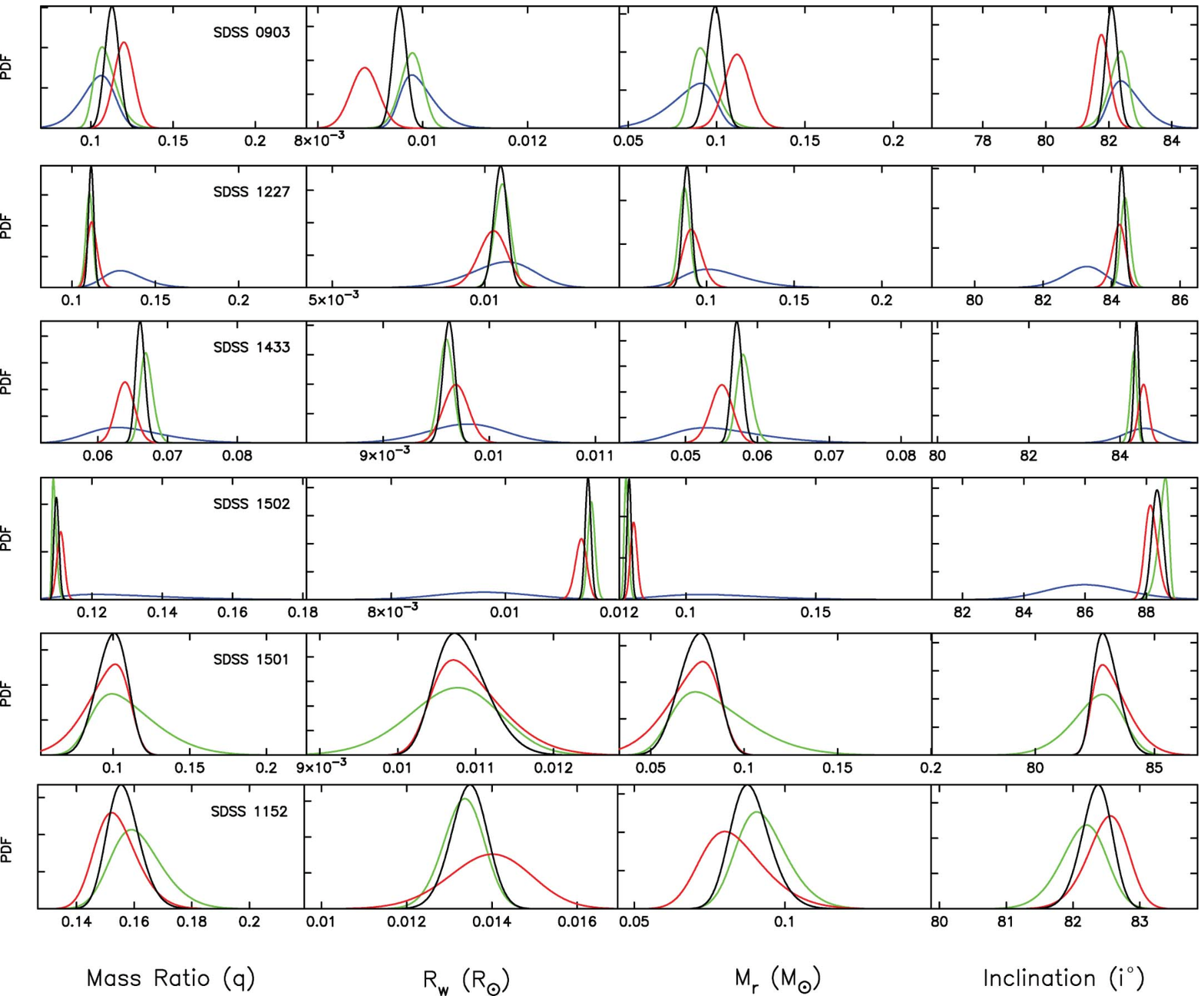}
\end{center}
\contcaption{}
\end{figure*}

\begin{table*}
\vbox to220mm{\vfil Table 3 (system parameters) goes here (landscape). See separate file (table3params).
\caption{}
\label{table:system_params}
\vfil}
\end{table*}

\subsection{Notes on individual systems}
\label{sec:notes}

\subsubsection{CTCV 1300}
\label{sec:ctcv1300_2010}

As noted in section \ref{sec:pme}, the two eclipses of CTCV 1300 were modelled individually due to the heavy pre-eclipse
flickering present in each light curve. The $u'g'r'$ eclipses from the night of 2007 June 10 gave consistent results, as 
did the $u'g'i'$ eclipses from the night of 2007 June 13. However, results from the individual nights were not consistent 
with each other, and we are presented with two distinct solutions which are the result of heavy pre-eclipse flickering 
altering the shape of the bright spot ingress feature. This in turn gives two very different values for the mass ratio, 
$q$. Encouragingly we find our white dwarf masses and radii are consistent between nights. To derive the final system 
parameters we use the PDFs as outlined in section \ref{sec:pme} and then take an average of the solution from each 
night. The error is taken as the standard deviation between the two values.  

\begin{figure}
\centering
\includegraphics[scale=1.10,trim=0 0 0 0]{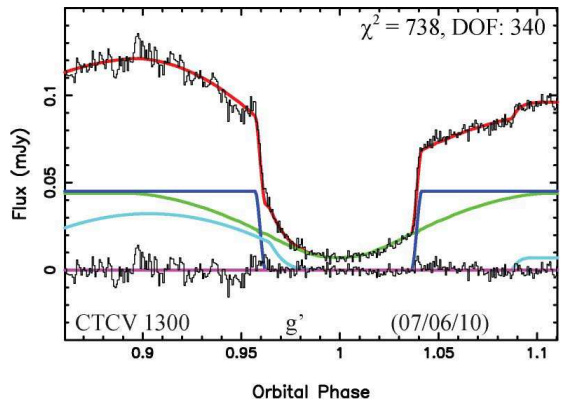}
\caption{Our June 2010 $g'$ band eclipse of CTCV 1300, together with a modified model found using the 2007 data.
Starting from an average of the two 2007 models, we ran a downhill simplex fit varying all parameters bar $q$, 
$\Delta\phi$, $R_{w}/a$ and $U_{w}$.}
\label{fig:ctcv1300_2010}
\end{figure}

In Fig. \ref{fig:ctcv1300_2010} we see a $g'$-band eclipse from June 2010 together with a modified version of the 
model obtained from the 2007 dataset. This new model is found by starting from an average of the two 2007 fits and 
using a downhill simplex method to vary all parameters bar $q$, $\Delta\phi$, $R_{w}/a$ and $U_{w}$. These parameters
should not change with time, and so the simplex fit will confirm if our bright spot positions and white dwarf dwarf 
radius are correct, and thus if our system parameters are reliable. We use a simplex method for two reasons; firstly 
since we are performing a consistency check and are not extracting system parameters from the fitting process we do 
not require a full MCMC analysis. Secondly, the bright spot flux appears to have reduced significantly, and the 
strength of the bright spot ingress feature means that we cannot constrain a full fit using the MCMC model used previously. 
The model confirms the bright spot flux has decreased considerably, although an orbital hump is still visible. The white 
dwarf flux remains almost unchanged, although the disc appears brighter. The fit to the data is good, indicating that 
the models derived from the 2007 data (and used to derive our system parameters) are reliable.

\subsubsection{CTCV 2354}
\label{sec:ctcv2354_2010}

\begin{figure*}
\centering
\includegraphics[scale=0.8,trim=0 0 0 0]{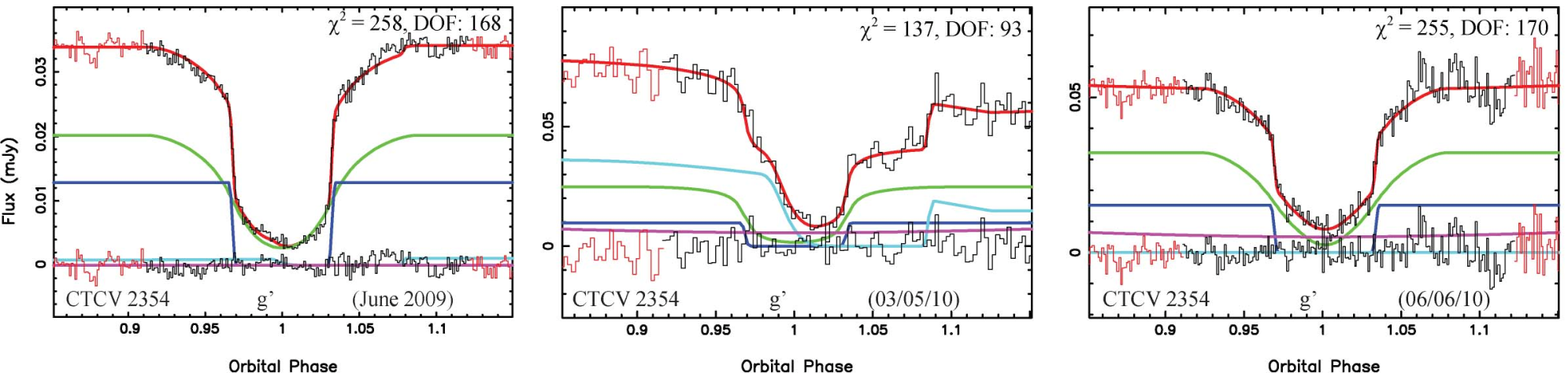}
\caption{Our $g'$ band observations of CTCV 2354 from 2009 (left), 2010 May 3 (centre) and 2010 June 6 (right), 
fit with a modified version of the model obtained from the 2007 data as described in section \ref{sec:ctcv2354_2010}.}
\label{figure:ctcv2354_2009_2010}
\end{figure*}

In section \ref{sec:2354} we noted that the shape of the light curve in Fig. \ref{figure:lightcurves} indicated 
possible bright spot egress features around phases 1.060 and 1.080. Fig. \ref{fig:eclipses} shows that our model has 
fit the bright spot egress feature at phase 1.080. Given the strength and shape of the bright spot features and general 
scatter present in the light curve we cannot be certain if the bright spot positions have been correctly identified by our 
model, and thus there is some element of doubt as to the value obtained for our mass ratio and thus donor mass. 
It is at this point we draw the readers attention to our 2010 data, shown in Fig. \ref{figure:ctcv2354_2009_2010}.  
The eclipse dated 2010 May 3 (centre panel) shows clear bright spot ingress and egress features, with a clear 
orbital hump visible from phases 0.70--0.95. The system is much brighter in this state than the 2007 data previously
modelled, in part due to a dramatic increase in bright spot flux. As with CTCV 1300 we carried out a downhill
simplex method to vary all parameters bar $q$, $\Delta\phi$, $R_{w}/a$ and $U_{w}$. Our fit to the eclipse of May 3 
is especially pleasing, as its excellent agreement with the light curve confirms that our 2007 model correctly identified 
the bright spot egress feature and thus the mass ratio obtained is reliable. 

Fig. \ref{figure:ctcv2354_2009_2010} also shows a single eclipse observed just one month later (June 2010, right panel), 
and six eclipses averaged together from June 2009 (left panel). Both of these datasets are fit with a downhill simplex model 
as above. The June 2009 and June 2010 datasets are in stark contrast to the May 2010 data, with the bright spot features 
appearing extremely faint (2009), or seemingly non-existent (June 2010). The disc flux in the June 2010 data appears to 
have increased significantly, giving rise to a distinct ``u'' shape. The rapid change in bright spot and disc light curves 
over such short (1 month - 1 year) time scales suggests that the disc is highly unstable.

\subsubsection{DV UMa}

Our donor mass derived for DV UMa has increased by $6.1\sigma$ ($\Delta$M=0.039$M_{\odot}$) from that published by Feline et 
al. (2004b). \nocite{feline2004b} Close inspection of the original fit reveals that the bright spot features are fit 
poorly by the old bright spot model. This arises because the old bright spot model could not describe the complex bright 
spot profile present and an innacurate value of the mass ratio is found as a result. Our new bright spot model is much 
better in this respect, and is able to take into account a wider variety of geometric effects and orientations. Given 
that our white dwarf radius is consistent with that of Feline et al. (2004b), this seems the most likely cause of such 
a large change. It is worth noting that our new donor masses for both DV UMa and XZ Eri, are both consistent with the 
masses obtained by Feline et al. (2004b) using the derivative method, which, unlike our parameterised model, does not 
make any attempt to recreate the bright spot eclipse profile (e.g. Wood et al. 1986; Horne et al. 1994; Feline et al. 
2004a; Feline et al. 2004b). \nocite{wood1986, horne1994, feline2004a, feline2004b}

\subsubsection{SDSS 1502}

Our new fits to SDSS 1502 decrease the donor mass by $2.9\sigma$ ($\Delta$$M_{r}=0.012M_{\odot}$) from that of 
\citet{littlefair2008}. Our mass 
ratio and inclination are consistent with those of \citet{littlefair2008}, however 
our white dwarf radius, $R_{w}$, has increased by $13$ percent ($3.4\sigma$). We believe the primary reason for 
this change was that the original fit was heavily binned, and thus more susceptible to the bug outlined in 
section \ref{sec:pme}.

\subsubsection{SDSS 1501}
\label{sec:SDSS1501}

The most important change of all of our re-modelled systems is for that of SDSS 1501. Whilst our donor mass 
has only increased by $1.9\sigma$ from that of \citet{littlefair2008}, we note that our uncertainties are large
($\sigma$$M_{r}=0.010M_{\odot}$) and the mass difference is large enough to take this system from being a 
post-period-bounce system, to a pre-period-bounce system. 
Although our errors do not formally rule out the possibility of SDSS 1501 being post-period-bounce, the donor's position with 
respect to the evolutionary tracks shown in Fig. \ref{figure:models} strongly favours that of a pre-period-bounce 
system. Such a change arises from a difference in bright spot positions between our model and that of \citet{littlefair2008}, 
which in turn affect the mass ratio obtained. The data used by \citet{littlefair2008} shows a very weak bright spot 
ingress feature. With the improvements made to the modelling process resulting from the introduction of MCMC, it is
clear that the 2006 data used by \citet{littlefair2008} does not constrain the mass ratio, $q$, tightly enough. 
In contrast, the 2004 data shows much clearer bright spot features, and we therefore favour this over the 2006 data 
as discussed in section \ref{sec:pme}.

\section{Discussion}
\label{sec:discussion}

\subsection{White dwarf masses}
Population studies by \citet[][]{willems2005} predict that between 40 and 80 percent of CVs are born with He-core white 
dwarfs ($M_{w} \lesssim 0.50 M_{\odot}$) and therefore He-core white dwarfs (He-WDs) are expected to be common amongst CV 
primaries. It is surprising then that out of our sample of 14 systems, we observe no He-WDs. Of all of our objects, SDSS 
1152 is found to have to the lowest white dwarf mass with $M_{w} = 0.560\pm0.028$. The mass distribution of \citet{kepler2007} 
for SDSS white dwarfs suggests He-WDs have a typical mass of $\sim$0.38$M_{\odot}$. The most massive He-WDs are thought to 
form from single RGB stars, which due to extreme mass loss are able to avoid the Helium flash; \citet{dcruz1996} consider 
models with a range of mass loss rates on the RGB and manage to produce He-WDs with masses up to $\sim$0.48$M_{\odot}$. It 
is likely that this represents an upper limit to the mass of He-WDs and hence SDSS 1152 is too massive to be a viable 
candidate for a He-core white dwarf.

We find that our white dwarf masses are not only too massive to be He-WDs, but are also well in excess of the average 
mass for single DA white dwarfs. Using the same method as \citet{knigge2006}, we calculated the average white dwarf mass 
of our entire sample to be $M_{w} = 0.81\pm0.04 M_{\odot}$, with an intrinsic scatter of $0.13 M_{\odot}$. 
In comparison, \citet{liebert2005} find the mean mass of DA white dwarfs to be $M_{w}\sim$0.603 $M_{\odot}$, 
while \citet{kepler2007} find a mean mass of $M_{w} = 0.593\pm0.016 M_{\odot}$.

\begin{figure*}
\centering
\includegraphics[scale=0.6, angle=-90]{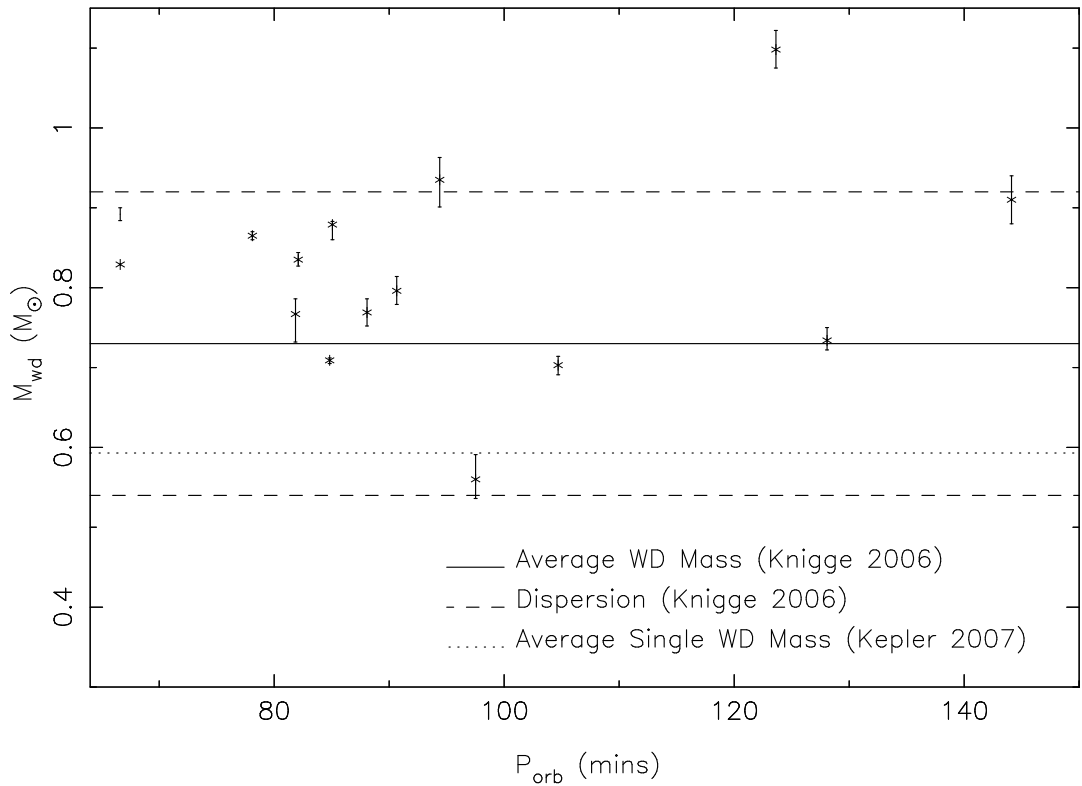}
\caption{White dwarf mass as a function of orbital period. The mean white dwarf mass for systems below the period gap, 
as found by \citet{knigge2006} is shown with a solid line, along with the associated intrinsic scatter (dashed line).
The mean white dwarf mass in single stars as found by \citet{kepler2007} is shown by a dotted line.}
\label{figure:wd_masses}
\end{figure*}

Our study thus supports previous findings \citep[e.g.][]{warner1973, warner1976, ritter1976, ritter1985, robinson1976, 
smith1998, knigge2006} that white dwarfs in CVs are on average 
much higher in mass than single field stars. Like \citet{littlefair2008}, we compare our masses to the average mass of 
$M_{w} = 0.73\pm0.05 M_{\odot}$ for white dwarfs in CVs below the period gap \citep{knigge2006} and find that our white 
dwarf masses are generally much higher. This is especially so for systems $P_{orb} \leq 95$ mins, where we find a mean 
white dwarf mass of $M_{w} = 0.83\pm0.02 M_{\odot}$, with an intrinsic scatter of $0.07 M_{\odot}$. Some our white dwarf 
masses are revised, and have accordingly moved downwards in mass compared to \citet{littlefair2008}, but as Fig. 
\ref{figure:wd_masses} shows, for $P_{orb} \leq 95$ mins, 8 out of 9 systems are more massive than the average found by 
\citet{knigge2006}, with SDSS 1502 the only exception. On the same plot, we also plot the dispersion of masses as found 
by \citet{knigge2006} and the mean mass of single SDSS white dwarfs found by \citet{kepler2007}. Note that the 
\citet{knigge2006} sample contains three systems also included in our study: OU Vir, DV UMa and XZ Eri, using the 
old mass determinations of Feline et al. (2004a, 2004b). We see most of our masses are within the dispersion found, 
indicating that no individual white dwarf mass is unusual. However 8 out of 9 systems above average does seem anomalously 
high, considering that if we model the white dwarf masses as a Gaussian distribution, the probability of such an occurrence 
is less than 2 percent (independent of the actual mean or variance). Such difference between our sample and that of 
\citet{knigge2006} is concerning, and it is therefore desirable to consider the selection effects, considering the 
majority of our short period systems are all SDSS objects \citep{szkody2004, szkody2005, szkody2006, szkody2007}. 

The majority of SDSS CVs found are generally rejected quasar candidates with a limiting magnitude of $g'=19-20$, 
and are initially selected for follow up on the basis of $u'$-$g'$ colour cuts (see \citet{gaensicke2009}
for a more in depth description). \citet{littlefair2008} show that systems with $M_{w} \geq0.50 M_{\odot}$ are blue enough
to pass the SDSS colour cuts and conclude that selection effects such as these are unlikely to explain the high 
mass bias of our white dwarf sample. However, the majority of the systems we have studied are close ($g'$$\sim17.5-19.5$)
to the $g'=19$ limit of the SDSS survey. This raises the possibility that the SDSS sample only finds the brightest of the 
short period CVs. \citet{ritter1986} have shown that CVs with high mass white dwarfs are brighter than their low mass 
counterparts. This suggests there maybe some bias towards high mass dwarfs. However, this conclusion is not appropriate
to the systems studied here: \citet{ritter1986} only consider the effects of accretion luminosity whereas in most of 
our systems the white dwarf considerably outshines the accretion disc. 

\citet{zorotovic2011} consider selection effects in white dwarf dominated SDSS systems with a variety of different 
mass transfer rates and conclude there is actually bias {\it against} high mass white dwarfs. They find a 0.90$M_{\odot}$ 
white dwarf is approximately 0.15 magnitudes fainter than a 0.75$M_{\odot}$ white dwarf, which corresponds to a decreased 
detection efficiency of $\sim20\%$. This suggests that our finding of high mass white dwarfs at short orbital periods is 
not due to selection effects, and is in fact a true representation of the intrinsic mass distribution of CVs. However, 
this analysis only considers white dwarf luminosity; in some of our systems the bright spot features are 
prominent, and contribute significantly to the overall flux of the system. It remains possible that the finding of 
very high white dwarf masses for our short period CVs is due to selection effects. However, a full and thorough quantification 
of any bias would require detailed calculations of the luminosity of white-dwarf dominated systems (including bright 
spot emission) plus an investigation of the selection effects in the SDSS sample. Such an analysis is beyond the
scope of this paper.

Our results have important consequences for the modelling of nova outbursts and their impact on the long-term evolution 
on CVs. Typical calculations show that the mass of the white dwarf decreases by between 1 and 5 percent per 1000 nova cycles 
\citep[e.g.][]{yaron2005, epelstain2007}. The dominance of high-mass white dwarfs in our sample of short period systems 
suggests that any white dwarf erosion due to nova explosions must be minimal, or that not all of the accreted matter is 
ejected during nova ignition, resulting in the white dwarf mass {\it increasing} over time. This could, in principle 
enable the white dwarfs in cataclysmic variables to grow in mass until they reach the Chandrasekhar limit.

\subsection{Period bounce}

Population synthesis models for cataclysmic variables \citep[e.g.][]{kolb1993, willems2005} all predict that large 
numbers of the CV population ($\sim$15 - 70 percent) have evolved past the period minimum. This has always been in 
stark contrast to observations, possibly in part due to selection effects \citep[e.g.][]{littlefair2003}.

\citet{littlefair2006b, littlefair2007, littlefair2008} identified four systems (SDSS 1035, SDSS 1507, SDSS 1433, 
SDSS 1501) with donors below the sub-stellar limit, three of which are likely to be post-period-bounce CVs (SDSS 
1035, SDSS 1501 and SDSS 1433). Our subsequent re-analysis gives three systems (SDSS 1035, SDSS 1507, SDSS 1433) with
donors below the sub-stellar limit, two of which (for reasons outlined below) we believe are post-period-bounce 
CVs (SDSS 1035 and SDSS 1433). SDSS 1501, which no longer features as a post-period-bounce system, is discussed in 
greater detail in section \ref{sec:SDSS1501}. 

\citet{sirotkin2010} claim that SDSS 1433 cannot be considered a post-period-bounce object since the mass transfer 
rates and donor star temperatures implied are too high. The mass transfer rate is 
found using an estimate of the white dwarf temperature \citep{townsley2009}, while the donor star temperature 
is inferred using a semi-empirical relationship that is also dependent on the white dwarf temperature. The white 
dwarf temperature used by \citet{sirotkin2010} is that derived by \citet{littlefair2008} from model fitting.
We believe, at least in this case, that using $\dot{M}$ and $T_{2}$ is an unreliable test of the evolutionary status 
of CVs donors, since accurate determinations of the white dwarf temperature are difficult to obtain. Of all the 
system parameters we have derived, the white dwarf temperatures are the least well constrained, and this does not take 
into account systematic errors. Since the white dwarf temperature 
is found using the flux from just three colours, and our model does not include all possible sources of luminosity 
(e.g. a boundary layer), there is a good chance our white dwarf temperatures are affected by systematic errors at some level, 
as discussed in section \ref{sec:pme}. Instead, we focus on the donor star mass, $M_{r}$.

If the angular momentum loss rate is similar for systems with identical system parameters, we expect all CVs to 
follow very similar evolutionary tracks 
with a single locus in the mass-period relationship (and by analogy, mass-radius relationship) for CV donors, as 
shown in Fig. \ref{figure:models}. The empirical donor star mass-radius relationship derived by \citet{knigge2006} 
shows that a single evolutionary track does very well at describing the observed $M_{2}-P_{orb}$ relationship, although
the shape of that relationship is poorly constrained at low masses. A single evolutionary path also explains the 
presence of the ``period spike'', a long sought after feature in the orbital period distribution recently identified 
by \citet{gaensicke2009}. We therefore expect there to be a unique donor mass corresponding to the minimum orbital 
period, below which an object becomes a period-bouncer. The exact mass at which this occurs is very uncertain, and 
does not necessarily correspond to the sub-stellar limit \citep{patterson2009}. From the empirical work of \citet{knigge2006}, 
the best estimate for $M_{bounce}$ is $M_{r} = 0.063 \pm 0.009 M_{\odot}$. Three of our systems (SDSS 1035, SDSS 1433 
and SDSS 1507) fall well below this value, although SDSS 1507 is an unusual system, and is discussed in the following 
section. As in \citet{littlefair2008}, we do not include it in our sample of post-period minimum CVs. We therefore have two 
strong candidates for post-period minimum CVs (SDSS 1035 and SDSS 1433) from our total sample of 
14 CVs (nine of which are SDSS systems). From this, we estimate that $14 \pm 7$ percent of all CVs below the period
gap, and $22 \pm 11$ percent of all short period CVs ($P_{orb} \leq 95$ mins) have evolved 
past the period minimum. These findings are consistent, albeit to a crude approximation given our small sample of objects, 
with current population synthesis models. Since all of our short period systems are SDSS CVs, we cannot rule out selection 
effects, but \citet{gaensicke2009} have shown that the number of period minimum CVs found within the SDSS is broadly 
consistent with other surveys, allowing for normalisation of survey volumes.

\subsection{SDSS 1507}

The orbital period of SDSS 1507 is far below the well-defined period minimum and thus the nature of this system is of
great interest to theorists and observers. It is possible that this system represents the true orbital period minimum 
as predicted by \citet{kolb1999}. However, if this is indeed the case, we would expect a large number of systems between 
orbital periods of 67 minutes, and 83 minutes where the period spike is observed \citep{gaensicke2009}. These systems are 
not observed, and hence it is likely that some other mechanism is responsible. \citet{littlefair2007} speculate that this 
system was either formed directly from a white dwarf/brown dwarf binary, while \citet{patterson2008} argue that the system 
could be a member of the halo. Both derive system parameters, and both obtain distance estimates to the system.

Our derived system parameters are consistent with those of \citet{littlefair2007} and \citet{patterson2008}, within
uncertainties. Our distance estimate is in excellent agreement with \citet{littlefair2007}, which is not suprising 
since we both calculate the distance using the same methods and dataset. However, our new distance estimate still 
places the system nearer than that of \citet{patterson2008}. \citet{patterson2008} obtain a lower limit to the
distance using parallax. The parallax value implies a distance $d > 175$ pc, which taken alone,
is consistent with our estimate of $d=168\pm12$ pc. Patterson combines his parallax with a range of other
observational constraints using Bayesian methods to 
yield a final distance estimate of $d=230\pm40$ pc. If our distance of $d=168\pm12$ pc is nearer the true distance, 
then combining with Patterson's proper motion measurement of $0.16 ''/yr$ yields a transverse velocity of $d=128\pm9$ 
kms$^{-1}$. This lower transverse velocity is still very much an outlier in the distribution of 354 CVs shown in Fig.1 of 
\citet{patterson2008}. Therefore, regardless of which distance is correct, the proper motion of SDSS 1507 still 
supports halo membership.

\subsection{Exploring the standard model of CV evolution}
\begin{figure*}
\centering
\includegraphics[scale=0.6, angle=0]{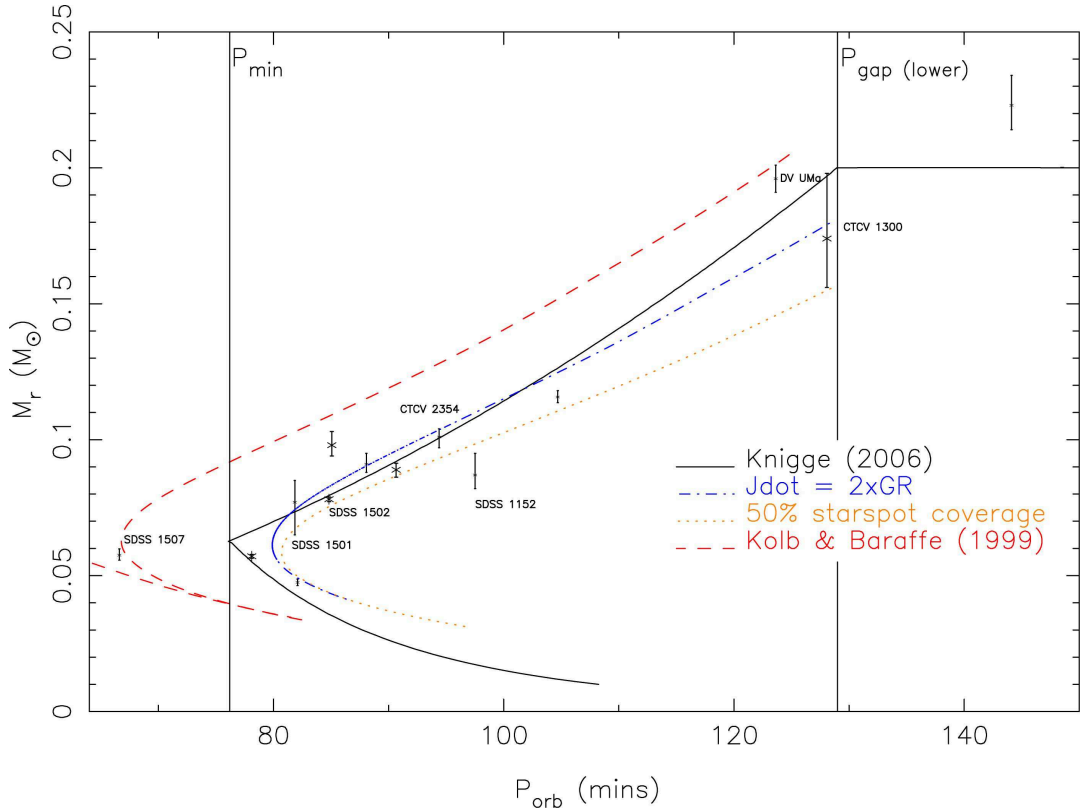}
\caption{The $M_{2}-P_{orb}$ relationship for our dataset. Mass determinations for all systems using {\sc ultracam} 
data are included. Our three new systems, in addition to other objects of particular interest are labelled. 
The evolutionary models of \citet{kolb1999} calculated with different mass-transfer rates are shown with red 
(dashed) and blue (dot-dashed) lines. A model with 50 percent star spot coverage on the surface of the donor  
is shown with an orange (dotted) line. The solid (black) line shows the empirical mass-radius relationship as 
found by \citet{knigge2006}. The position of the period minimum and period gap, as found by \citet{knigge2006}, 
are also shown.}
\label{figure:models}
\end{figure*}

Fig. \ref{figure:models} shows the evolutionary models of \citet{kolb1999} calculated with enhanced mass-transfer 
rates. Also shown is a model with 50 percent star spot coverage on the surface of the donor. Positions of the period 
minimum, and period gap as found by \citet{knigge2006} are also shown. Mass determinations for all systems presented
here are included. We see that the standard theoretical models are a poor fit to the data. For a given mass, the models of 
\citet{kolb1999} significantly underestimate the orbital period, and thus the donor radii.

Models with enhanced mass transfer rates and star spot coverage do rather better at reproducing the observed donor
masses, although the general scatter of short period systems makes choosing between these difficult. This is in
line with the conclusions of \citet{littlefair2008}. The models begin to diverge significantly at orbital 
periods greater than 100 minutes. Unfortunately, in this regime there are few systems with precisely known donor masses. 
Clearly, we require more mass determinations for systems with orbital periods between 100 and 130 minutes.

\section{Conclusions}
\label{sec:conclusion}

We present high-speed, three-colour photometry of a sample of 14 eclipsing CVs. Of these CVs, nine are short period 
($P_{orb} \leq 95$ minutes), and one is within the period gap. For each of the 14 objects we determine the system 
parameters by fitting a physical model of the binary to the observed light curve by $\chi^{2}$ minimisation. We find 
that two of our nine short period systems appear to have evolved past the period minimum, and thus 
supports various assertions that between 15 and 70 per cent of the CV population has evolved past the orbital period
minimum. The donor star masses and radii are not consistent with model predictions, with the majority of donor stars 
being $\sim$10 per cent larger than predicted. Our derived masses and radii show that this can explained by either 
enhancing themass transfer rate or modifying the stellar physics of the donor star to take into account star spot coverage. 
Unfortunately, we still lack enough precise donor masses between orbital periods of 100 and 130 minutes to choose between 
these alternatives.

Finally, we find that the white dwarfs in our sample show a strong tendency towards high masses. The high mass dominance
implies that the white dwarfs in CVs are not significantly eroded by nova outbursts, and may actually increase over several nova 
cycles. We find no evidence for He-core white dwarfs within our sample, despite predictions that between 40 and 80 percent
of short period CVs should contain He-core white dwarfs.

\section{Acknowledgements}

We would like to thank our referee, Joe Patterson for his useful comments. We also
thank Christian Knigge for useful discussions on white dwarf bias and selection effects. CDJS acknowledges 
the support of an STFC PhD. SPL acknowledges the support of an RCUK Fellowship. CMC and TRM are supported under grant
ST/F002599/1 from the Science and Technology Facilities Council (STFC). {\sc ultracam} and 
{\sc ultraspec} are supported by STFC grant ST/G003092/1. This research has made use of NASA's Astrophysics Data System 
Bibliographic Services. This article is based on observations made with {\sc ultracam} mounted on the Isaac Newton Group's 
WHT, and {\sc ultracam} and {\sc ultraspec} mounted on the European Southern Observatory's NTT and VLT telescopes.

\bibliographystyle{mn2e}
\bibliography{myrefs}
\label{lastpage}
\end{document}